\documentclass[11pt]{article}
\usepackage[dvips]{graphicx}
\usepackage{cite}
\usepackage{amsmath}
\usepackage{amsfonts}
\renewcommand{\vec}[1]{\boldsymbol{#1}}
\newcommand{\myhat}{\hat}
\newcommand{\ad}{{\rm ad}}
\newcommand{\TB}[0]{T_B}
\newcommand{\omm}{{ \textstyle\frac{\omega_Bt}{2}}}
\newcommand{\ON}{\myhat{N}}
\newcommand{\OK}{\myhat{K}}
\newcommand{\OKA}{\myhat K^\dagger}
\newcommand{\OU}{\myhat{U}}
\newcommand{\OI}{{\myhat I}}
\newcommand{\OJ}{{\myhat J}}
\newcommand{\OJA}{\myhat J^\dagger}
\newcommand{\M}{{\mathbb{M}}}
\newcommand{\LL}{{\mathcal{L}}}
\newcommand{\ev}[1]{\langle #1 \rangle}
\newcommand{\RR}{{\mathcal I}}
\renewcommand{\SS}{{\mathcal S}}
\newcommand{\sds}{\subset\hspace{-12pt}+}
\newcommand{\rmi}{{\rm i}}
\newcommand{\rmd}{{\rm d}}
\newcommand{\rme}{{\rm e}}
\newcommand{\Z}{\mathbb{Z}}
\newcommand{\N}{\mathbb{N}}
\newcommand{\R}{\mathbb{R}}
\newcommand{\ket}[1]{% arg1: Eintrag im ket-Vektor
	  | #1 \rangle}

\newcommand{\scal}[2]{% arg1: Eintrag im bra-Vektor
			\ensuremath{\langle #1 | #2 \rangle}}
\newcommand{\ketbra}[2]{
			\ensuremath{|#1\rangle \langle  #2|}}
\skip\footins 15pt plus 4pt minus 2pt
\begin{document}
\title{Two-dimensional Bloch oscillations:\\ A Lie-algebraic approach}
\author{S. Mossmann, A. Schulze, D. Witthaut, H. J. Korsch\\
FB Physik, University of Kaiserslautern\\ D-67653
Kaiserslautern\\ Germany}
\maketitle
\noindent{\it Email: korsch@physik.uni-kl.de}\vspace{0.5cm}\\
{\it ABSTRACT}\\
A Lie-algebraic approach successfully used to describe one-dimensional Bloch oscillations in a tight-binding approximation is extended to two dimensions. This extension has the same algebraic structure as the one-dimensional case while the dynamics shows a much richer behaviour. The Bloch oscillations are discussed using analytical expressions for expectation values and widths of the operators of the algebra. It is shown under which conditions the oscillations survive in two dimensions and the centre of mass of a wave packet shows a Lissajous like motion.
In contrast to the one-dimensional case, a wavepacket shows systematic dispersion that depends on the direction of the field and the dispersion relation of the field free system.\vspace{0.5cm}\vspace{0.5cm}

\section{Introduction}
The quantum dynamics of a particle in a tilted periodic potential, a so called Wannier-Stark system, has been the subject of investigation from the beginning of quantum mechanics \cite{Bloc28,Zene32,Rauh74}. In the early years the motivation was to understand the properties of an electric current in solid crystals while in recent years the realisation of semiconductor superlattices and the experimental progress in manipulating cold atoms resp. Bose-Einstein condensates in optical lattices has motivated new theoretical studies in this field (see \cite{04bloch1d} and the references given there).
Bloch oscillations are one of the striking phenomena in such systems and an example for a counter-intuitive behaviour in quantum mechanics \cite{03TBalg,04bloch1d,Naza89,Bouc95,03bloch2d,04bloch2d,Dmit01,Dmit02a}.
While \cite{Naza89,Bouc95,04bloch1d,03TBalg} discuss the one-dimensional case analytically in a single-band tight-binding model, in \cite{03bloch2d,04bloch2d,Dmit01,Dmit02a} two-dimensional Bloch oscillations are considered, however, only partially in an analytical treatment. \\
The aim of the present paper is to supplement the discussion of Bloch oscillations in two-dimensional lattices by
an analytic treatment within the tight-binding model.
The Lie-algebraic approach to the tight-binding model in
\cite{03TBalg} is extended to two space dimensions using the techniques of algebraic
structure theory.
Within this model, not only the nearest neighbour couplings along the axes (the so called von-Neumann neighbourhood) is taken into account, but also the interactions along the diagonals (Moore neighbourhood) will be considered and explicit expressions for the expectation values of the position operators will be derived. It is shown that Bloch oscillations in two dimensions occur only for special initial conditions and that typically the oscillation is superposed by a directed motion orthogonal to the field. On the basis of this examination the results can be extended to the single-band model where all couplings within one band are taken into account.

A two-dimensional Wannier-Stark system can be described by a Hamiltonian with a periodic potential $V$ and an additional linear potential:
\begin{equation}\label{WSH}
\myhat H = \frac{\myhat{\vec{p}}^2}{2}+ V(\myhat{\vec{r}}) +\vec{F}\cdot\myhat{\vec{r}}\quad\text{with}\;V(\vec{r}+\vec{r}_{\vec{n}}) = V(\vec{r})
\end{equation}
and $\vec{r_n} = n_1\vec{e}_1+n_2\vec{e}_2$, $n_i\in\Z$. For simplicity the period in both directions is set equal to one. The
force $\vec{F} = \vec{F}_0+\vec{F}_\omega(t)$ consists of a static part and a part with an arbitrary time dependence.
A convenient approximation is to expand the Hamiltonian in a basis of Wannier-functions $\ket{\vec{n},l}$ which are localised at site $\vec{n}= (n_1,n_2)$ of the periodic potential. The index $l$ labels the different bands. Taking into account only a single band and the coupling between adjacent sites, the Hamiltonian for the one-dimensional system can be written as
\begin{equation}
\myhat{H} = \frac{\Delta}{4}\sum_{m=-\infty}^{+\infty}
\Bigl(\ketbra{m}{m+1}+\ketbra{m}{m-1}\,\Bigr)+
F\sum_{m=-\infty}^{+\infty} m\, \ketbra{m}{m}\,,
\end{equation}
where $\ket{m}$ are the one-dimensional Wannier-states and $\Delta$ is the band width of the energy dispersion relation for the field-free system.
An alternative tight-binding approach using an expansion in the eigenstates of the Wannier-Stark Hamiltonian~\eqref{WSH} for a constant field can be found in \cite{Thom04}.

The tight-binding Hamiltonian can be easily extended to two
dimensions using the two-dimensional Wannier-states $\ket{m,n}$ with $\scal{m,n}{m',n'} = \delta_{m',m}\delta_{n',n}$. Then, the
Hamiltonian reads
\begin{align}
\myhat H &= \frac{\Delta_1}{4}\sum_{m,n=-\infty}^{+\infty}
\Bigl(\ketbra{m,n}{m+1,n}+\ketbra{m,n}{m-1,n}\Bigr)\nonumber\\
  &\quad+\frac{\Delta_2}{4}\sum_{m,n=-\infty}^{+\infty}\Bigl(
\ketbra{m,n}{m,n+1}+\ketbra{m,n}{m,n-1}\,\Bigr)\nonumber\\
  &\quad+\sum_{m,n=-\infty}^{+\infty} (F_1 m+F_2 n)\, \ketbra{m,n}{m,n}\,.
\end{align}
Introducing the operators
\begin{equation}\label{la-2dOpdef1}
\ON_1 =\! \sum_{m,n=-\infty}^{+\infty} m\, \ketbra{m,n}{m,n}\,,\quad
\ON_2 = \!\sum_{m,n=-\infty}^{+\infty} n\, \ketbra{m,n}{m,n}
\end{equation}
and
\begin{equation}\label{la-2dOpdef2}
\OK_1 =\! \sum_{m,n=-\infty}^{+\infty}  \ketbra{m,n}{m+1,n}\,,\quad
\OK_2 =\! \sum_{m,n=-\infty}^{+\infty}  \ketbra{m,n}{m,n+1}\,
\end{equation}
together with their adjoint operators $\OKA_1$ and $\OKA_2$
the notation can be simplified:
\begin{equation}
\myhat H = \frac{\Delta_1}{4}(\OK_1+\OK_1^\dagger)
+\frac{\Delta_2}{4}(\OK_2+\OK_2^\dagger)+F_1\ON_1+F_2\ON_2\,.
\end{equation}
The operators $\OK_1$ and $\OK_2$ are unitary, i.e. they fulfil the relation
\begin{equation}\label{la-2dunitary}
	\OK_j^{-1} = \OK_j^\dagger\quad \text{for}\quad j = 1,2
\end{equation}
and describe the coupling of adjacent sites along the two axes. The hermitian operators $\ON_1$ and $\ON_2$ can be interpreted as 'position' operators in direction 1 resp. 2.

For later convenience, the variable $\vec{f} = \vec{F}/\hbar$ will be
used instead of the field $\vec{F}=(F_1,F_2)$ and it is assumed that the components $f_{1}$ and $f_{2}$ of the field
$\vec{f}$ have a rational ratio, i.~e.
\begin{displaymath}
\frac{f_{1}}{f_{2}}= \frac{q}{r}\quad\text{with}\quad
q,r\in\Z\;\;\text{coprime} \,.
\end{displaymath}
Then, the field can be written as
\begin{equation}
\begin{pmatrix}f_{1}\\f_{2}\end{pmatrix} =
\frac{f}{\sqrt{q^2+r^2}}\begin{pmatrix}\,q\,\\\,r\,\end{pmatrix}
\end{equation}
with $f = |\vec{f}|$.
We will assume that the direction of the field is constant in time, whereas its magnitude ${f}$ can be time-dependent. For the example of Bloch oscillations studied below, the field is constant in time.
\section{The Lie-algebra}
The commutators of the operators $\ON_j$ and $\OK_j$ can be directly calculated using equations~\eqref{la-2dOpdef1} and \eqref{la-2dOpdef2} and the orthogonality of the Wannier-functions. The
non-vanishing commutators are
\begin{equation}
[\OK_j,\ON_j] = \OK_j\,,\quad[\OKA_j,\ON_j] = -\OKA_j\quad\text{for}\;j=1,2\,.
\end{equation}
So the operators $\{\ON_1,\OK_1,\OKA_1,\ON_2,\OK_2,\OKA_2\}$ form a Lie-algebra
\begin{equation}
\LL = \LL_1\oplus\LL_2
= \{\ON_1,\OK_1,\OKA_1\}\oplus
\{\ON_2,\OK_2,\OKA_2\}
\end{equation}
with $[\LL_1,\LL_2] = \{0\}$.
Since this algebra can be decoupled into two disjoint algebras which correspond to
the Lie-algebra of the one-dimensional tight-binding system \cite{03TBalg}, this
expansion is trivial. One can achieve further insight into the dynamics of the system
by
extending the  algebra by the operators
\begin{align}
\OK_1\OK_2 &= \sum_{m,n=-\infty}^{+\infty}  \ketbra{m,n}{m+1,n+1}\,,\\
\OK_1\OKA_2 &=\sum_{m,n=-\infty}^{+\infty}  \ketbra{m,n}{m+1,n-1}
\end{align}
together with the adjoints $\OKA_1\OKA_2$ and $\OKA_1\OK_2$. The ordering of these operators is arbitrary since they commute.
Adding these operators to the system takes couplings between
adjacent sites along the main diagonals into account and, from the algebraic point of view, couples both algebras $\LL_1$ and $\LL_2$\,.
The non-vanishing commutators for these operators are
\begin{eqnarray}
{[ \OK_1\OK_2,\ON_1 ]  }&= \phantom{-}\OK_1\OK_2\qquad
{[ \OK_1\OK_2,\ON_2 ]  }&= \phantom{-}\OK_1\OK_2 \nonumber\\
{[ \OKA_1\OKA_2,\ON_1 ]}&= -\OKA_1\OKA_2\qquad
{[ \OKA_1\OKA_2,\ON_2 ]}&= -\OKA_1\OKA_2\\
{[\OK_1\OKA_2,\ON_1]} &=\phantom{-} \OK_1\OKA_2\qquad
{[\OK_1\OKA_2,\ON_2]} &=-\OK_1\OKA_2\nonumber\\
{[\OKA_1\OK_2,\ON_1]} &=-\OKA_1\OK_2\qquad
{[\OKA_1\OK_2,\ON_2]} &= \phantom{-}\OKA_1\OK_2\,.\nonumber
\end{eqnarray}
Thus, this set of operators closes under commutation and forms a Lie-algebra
\begin{equation}
\LL=\{\ON_1,\,\OK_1,\,\OKA_1,\,\ON_2,\,\OK_2,\,\OKA_2,\,
\OK_1\OK_2,\,\OKA_1\OKA_2,\,\OK_1\OKA_2,\,\OKA_1\OK_2\} 
\end{equation}
with the corresponding Hamiltonian
\begin{align}\label{la-2dHam}
\myhat H &= \frac{\Delta_1}{4}(\OK_1 + \OKA_1)+\frac{\Delta_2}{4}(\OK_2 +
\OKA_2)\nonumber\\
& \quad+\frac{\Delta_3}{4}(\OK_1\OK_2 +\OKA_1\OKA_2+\OK_1\OKA_2+\OKA_1\OK_2)
+\hbar f_{1}\ON_1+\hbar f_{2}\ON_2\,.
\end{align}
The factors $\Delta_k$, $k = 1,\,2,\,3$ are the band widths of the dispersion relation in the direction of the axes ($\Delta_1$, $\Delta_2$) resp. the main diagonals ($\Delta_3$).
A discussion of the spectral properties of such two-dimensional tight-binding Hamiltonians taking into account various couplings can be found in \cite{02tb2d}.\\
For later convenience we introduce the notation
\begin{equation}\label{Kuvdef}
\OK_{u,v} = \OK_1^u\OK_2^v\quad\text{with}\quad u,v\in\Z\,,
\end{equation}
where the commutators read
\begin{equation}\label{Kuvcom}
[\OK_{u,v},\ON_j]= \epsilon_{u,v,j}\,\OK_{u,v}\quad\text{for}\;j = 1,2;\; u,v \in\Z
\end{equation}
with $\epsilon_{u,v,j} = u\,\delta_{j,1}+v\,\delta_{j,2}$\,.
Then the Lie-algebra can be written as (cf. equation~\eqref{la-2dunitary})
\begin{equation}\label{lauv}
\LL = \left\{\ON_1,\ON_2,\OK_{u,v}\Big|\,(u,v)\in \M\right\}
\end{equation}
with
\begin{equation}
\M = \{-1,0,1\}^2\setminus\{(0,0)\}\,,
\end{equation}
i.~e. the identity operator $\OI = \OK_{0,0}$ is not included in the algebra.
In section~\ref{sbm} the algebra will be extended by taking into account all possible operator products. Then, the Lie-algebra has  still the form \eqref{lauv}, but with $\M = \Z^2\setminus\{(0,0)\}$.

In this algebraic approach to the tight-binding model, all calculations can be done {\it without} using the explicit representations of the operators in \eqref{la-2dOpdef1} and \eqref{la-2dOpdef2}, only making use of the commutation relations in equation~\eqref{Kuvcom}.
\section{The time evolution operator}\label{teo}
As described in \cite{03TBalg,Wei63-1}, the Lie-algebra can be decomposed into
a semidirect sum, $\LL = \RR\sds\SS$, with
\begin{align}
\RR &= \left\{\OK_{u,v}\Big|(u,v)\in\M \right\}\quad\text{and}\\
\SS &= \{\ON_1,\,\ON_2\}\,.
\end{align}
Here, $\RR$ is an ideal of the Lie-algebra and $\SS$ is a subalgebra. The Hamiltonian can be written as a sum of operators, $\myhat H = \myhat H_S + \myhat H_I$, with $\myhat H_S\in\SS$ and $\myhat H_I\in\RR\,$:
\begin{align}
\myhat H_S  &= \hbar f_{1}\ON_1+\hbar f_{2}\ON_2\,,\\
\myhat H_I  &= \frac{\Delta_1}{4}(\OK_{1,0} + \OK_{-1,0})+\frac{\Delta_2}{4}(\OK_{0,1} +
\OK_{0,-1})\nonumber\\&\quad+\frac{\Delta_3}{4}(\OK_{1,1} +\OK_{-1,-1}+\OK_{1,-1}+\OK_{-1,1})\,.
\end{align}
In order to find the time evolution operator $\OU$ with
\begin{equation}
\rmi \hbar \,\frac{\rmd \myhat U}{\rmd t}=\myhat H\,\myhat U\,,
\end{equation}
one can make use of the algebraic structure, which allows a factorisation of the evolution operator as $\OU = \OU_S\, \OU_I$ with
\begin{equation}\label{la-2ddglU}
\rmi \hbar \,\frac{\rmd \myhat U_S}{\rmd t}=\myhat H_S\,\myhat
U_S \ ,  \quad
\rmi \hbar\,\frac{\rmd \myhat U_I}{\rmd t}=\big(\myhat
U_S^{-1}\myhat H_I\myhat
U_S\big)\ \myhat U_I\,
\end{equation}
(see \cite{88lie,Wei63-1,Wei63-2} for the details). Thus, the problem of calculating the time evolution operator is splitted into two parts. First one has to solve the differential equation for the subalgebra $\SS$ and afterwards solve the problem for $\RR$.

Before we begin with the derivation of the evolution operator, we introduce the so-called $\myhat\Gamma$-evolved operator $\myhat A$, defined by
\begin{align}\hspace{-0.1cm}
\rme^{z \,\ad \myhat \Gamma}\myhat A&=\rme^{z\myhat \Gamma}\myhat A\,\rme^{-z\myhat \Gamma}\nonumber
\\&=\myhat A+z\,\big[\,\myhat \Gamma,\myhat A \,\big]
+\frac{z^2}{2!}\,\big[\,\myhat \Gamma,\big[\,\myhat \Gamma,\myhat A \,\big] \,\big]
+\frac{z^3}{3!}\,\big[\,\myhat \Gamma,\,\big[\,\myhat \Gamma,\big[\,\myhat \Gamma,\myhat A \,\big] \,\big] \,\big]
+ \,\ldots\;.
\end{align}
with $\myhat\Gamma,\myhat A \in \LL$.\\
For the Lie-algebra $\LL$ these expression can be calculated directly using equation~\eqref{Kuvcom}. The non-trivial relations are
\begin{equation}\label{la-Gammaev}
\rme^{z\,\ad \OK_{u,v}}\,\ON_j = \ON_j + z\epsilon_{u,v,j}\OK_{u,v}\quad\text{and}\quad
\rme^{z\,\ad \ON_j}\,\OK_{u,v} = \rme^{-z\epsilon_{u,v,j}}\OK_{u,v}\,.
\end{equation}
After these preparations the first equation in \eqref{la-2ddglU} can be
solved and yields
\begin{equation}\label{USdef}
\OU_S(t) = \rme^{-\rmi(\eta_t^{(1)}\ON_1+\eta_t^{(2)}\ON_2)}
\end{equation}
with
\begin{equation}\label{eta1}
\begin{aligned}
\eta_t^{(1)} &= \int_0^t f_{1}(\tau) \,\rmd \tau
= \frac{q}{\sqrt{q^2+r^2}} \int_0^t f(\tau) \rmd \tau = q\, \eta_t\,,\\
\eta_t^{(2)} &= \int_0^t f_{2}(\tau)\, \rmd \tau
= r\, \eta_t\,,\\
\eta_t^{\phantom{(0)}} &= \frac{1}{\sqrt{q^2+r^2}}\int_0^t f(\tau) \,\rmd \tau\,.
\end{aligned}
\end{equation}
To solve the second differential equation in \eqref{la-2ddglU}, one has
to evaluate the right-hand side,
\begin{align}\hspace{-0.2cm}
\frac{1}{\hbar}\left(\OU_S^{-1}\myhat H_I\OU_S\right) &=
\frac{\Delta_1}{4\hbar}\left[\rme^{-\rmi q\eta_t}\OK_{1,0}+\rme^{+\rmi
q\eta_t}\OK_{-1,0} \right]
+\frac{\Delta_2}{4\hbar}\left[\rme^{-\rmi r\eta_t}\OK_{0,1}+\rme^{+\rmi
r\eta_t}\OK_{0,1}\right] \nonumber\\&\quad+
\frac{\Delta_3}{4\hbar}\left[\rme^{-\rmi(q+r)\eta_t}\OK_{1,1}
	+\rme^{+\rmi(q+r)\eta_t}\OK_{-1,-1}
	+\rme^{-\rmi(q-r)\eta_t}\OK_{1,-1}\right.\\&\quad
	+\left.\rme^{+\rmi(q-r)\eta_t}\OK_{-1,1}\right]\,,\nonumber
\end{align}
which is an element of $\RR$.
The solution of equation~\eqref{la-2ddglU} then reads
\begin{eqnarray}\label{URdef}
\OU_I(t) =& \exp\left\{-\rmi\bigl[\,
\chi_t^{(1,0)}\OK_{1,0}+{\chi_t^{(-1,0)}}\OK_{-1,0}+
\chi_t^{(0,1)}\OK_{0,1}+{\chi_t^{(0,-1)}}\OK_{0,-1}\right.\nonumber\\
&+\left.\chi_t^{(1,1)}\OK_{1,1}+\chi_t^{(-1,-1)}\OK_{-1,-1}+
\chi_t^{(1,-1)}\OK_{1,-1}+\chi_t^{(-1,1)}\OK_{-1,1}\bigr]\right\}
\end{eqnarray}
with
\begin{equation}
\chi_t^{(u,v)} =  g^{(u,v)}\int_0^t\,\rme^{-\rmi(uq+vr)\eta_\tau}\rmd
\tau\;,\quad (u,v) \in \M
\end{equation}
\text{and}
\begin{equation}
g^{(\pm1,0)}=\frac{\Delta_1}{4\hbar}\,,\quad g^{(0,\pm1)}=\frac{\Delta_2}{4\hbar}\,,\quad
g^{(\pm 1,\pm 1)}=g^{(\pm 1,\mp 1)}=\frac{\Delta_3}{4\hbar}\,.\nonumber
\end{equation}
The function $\eta_t$ is defined in equation~\eqref{eta1}. Note that the function  $\chi_t^{(-u,-v)}$ is the complex conjugate of $\chi_t^{(u,v)}$:
\begin{equation}
\chi_t^{(-u,-v)} = {\chi_t^{(u,v)}}^*\,.
\end{equation}
\subsection*{Time-independent fields:}
So far, the expressions are valid for an arbitrary, time dependent field $\vec{f}$.
For the time-independent case, $f = \text{const}$, the integration of \eqref{eta1} yields
\begin{equation}
\eta_t^{(1)} = \frac{qf}{\sqrt{q^2+r^2}}\,t\quad\text{and}\quad
\eta_t^{(2)} = \frac{rf}{\sqrt{q^2+r^2}}\,t\,.
\end{equation}
Defining the Bloch time $\TB$ for the two-dimensional system via
\begin{equation}
	\TB = \frac{2\pi}{f}\sqrt{q^2+r^2}\,
\end{equation}
and the Bloch frequency by $\omega_B = 2\pi/\TB$, the evolution
operator $\OU_S(t)$ can be written as
\begin{equation}
\OU_S(t) = \rme^{-\rmi\omega_B(q\ON_1+r\ON_2)t}\,.
\end{equation}
The functions
$\chi_t^{(u,v)}$ can also be evaluated directly,
\begin{align}\nonumber
\chi_t^{(u,v)} &= g^{(u,v)} \,\int\limits_0^t\rme^{-\rmi (uq+vr) \omega_B\tau}\rmd
\tau \\[2mm]&=
\begin{cases}
2\,\frac{ g^{(u,v)}}{(uq+vr)\omega_B}\sin\left((uq+vr)\omm\right)\,
\rme^{-\rmi (uq+vr)\omm}&\text{for}\!\!\!\quad uq+vr\neq0\\[2mm]
g^{(u,v)}t&\text{for}\!\!\!\quad uq+vr=0
\end{cases}\,.\label{chito}
\end{align}
Since both operators $\ON_j$ have an integer valued spectrum, the evolution operator over one Bloch period simplifies to
\begin{equation}
\OU(\TB) = \OU_S(\TB)\OU_I(\TB) =\OU_I(\TB)\,.
\end{equation}
Only if the field is {\it not} directed along one axis or a main diagonal, the operator $\OU_I(\TB)$ as defined in \eqref{URdef} is equal to the identity operator, $\OU_I(\TB) =  \OI$,  since then $uq+vr\neq0$ holds for all $(u,v)\in\M$ and therefore all functions $\chi_{\TB}^{(u,v)}$ vanish according to equation~\eqref{chito}. Otherwise, there exists one function $\chi_{\TB}^{(u,v)}$ with $uq+rv=0$, namely $\chi_{\TB}^{(r,-q)}$, that does not vanish according to $\chi_{\TB}^{(r,-q)}=g^{(r,-q)}\TB$.\\
In other words, if the coupling in the direction orthogonal to the field vanishes,
$g^{(r,-q)}=0$, resp. is not included into the model the evolution operator over one Bloch period is equal to the identity operator with the consequence that the dynamics of every initial state is periodic in time.

In addition, it should be pointed out that in the time-independent case the evolution operator can be directly written as
\begin{eqnarray}\nonumber
\OU(t) &=& \exp\{-\rmi \myhat H\,t/\hbar\}\\&=&\exp\Bigl\{-\rmi
\bigl[g^{(1,0)}(\OK_{1,0} + \OK_{-1,0})+g^{(1,0)}(\OK_{0,1} +
\OK_{0,-1})\nonumber\\
 &&+g^{(1,1)}(\OK_{1,1} +\OK_{-1,-1}+\OK_{1,-1}+\OK_{-1,1})
+f_{1}\ON_1+f_{2}\ON_2\bigr]\,t\Bigr\}\nonumber\\
&=& \exp\Bigl\{-\rmi \Bigl[\sum_{(u,v)\in\M}g^{(u,v)}\OK_{u,v}
+q\omega_B\ON_1+r\omega_B\ON_2\Bigr]\,t\Bigr\}
\end{eqnarray}
with $g^{(-u,-v)}=g^{(u,v)}$. The connection between this equation and the product form $\OU(t) = \OU_S(t) \OU_I(t)$ is given by the application of Taylor's theorem introduced by Sack\cite{Sack58}:
\begin{equation}
\exp\left\{ \kappa(
\myhat{A}+\lambda\myhat{B}
) \right\}
=\rme^{\kappa\myhat{A}}\exp\left\{
( \lambda\myhat{B}/a)
( 1-\rme^{-\kappa a})
\right\}
\end{equation}
where $\myhat{A}$ is a shift-operator,  $[ \myhat{A},\myhat{B}]\!=\!a\myhat{B}$, and $\lambda ,\,\kappa ,\,a$ ($a\neq0$) are complex numbers.
This relation can be directly extended to an arbitrary number of commuting shift-operators $\myhat B_\mu$ with $[\myhat{A},\myhat{B}_\mu]=a_\mu
\myhat{B_\mu}$ ($a_\mu\neq0$) and $[\myhat{B}_\mu,\myhat{B}_\nu]=0$:
\begin{equation}
\exp\bigl\{ \kappa\bigl(
\myhat{A}+\sum_\mu\lambda _\mu\myhat{B}_\mu
\bigl) \bigr\}
=\rme^{\kappa\myhat{A}}\,\prod\limits_\mu\exp\bigl\{
( \lambda_\mu\myhat{B}_\mu/a_\mu)
( 1-\rme^{-\kappa a_\mu})
\bigr\}\,.
\end{equation}
Identifying
\begin{equation}\nonumber
\myhat A =q\ON_1+r\ON_2\,,\;\,\myhat B_\mu = \OK^{(u,v)}\,,\;\,\kappa = -\rmi\omega_Bt\,,\;\,\lambda_\mu=g^{(u,v)}/\omega_B\,,\;\, a_\mu= -qu-rv\,,
\end{equation}
the evolution operator can be written in the product form
\begin{equation}
\OU(t)= \exp\Bigl\{-\rmi\omega_B(q\ON_1+r\ON_2)t\Bigr\}\,
\prod_{(u,v)\in\M}\exp\Bigl\{g^{(u,v)}\,\frac{\rme^{-\rmi(qu+rv)\omega_Bt}-1}{(qu+rv)\omega_B}\,\OK_{u,v}\Bigr\}\,,
\end{equation}
which is of course identical to $\OU(t)=\OU_S(t)\OU_I(t)$ given in equation~\eqref{USdef} and \eqref{URdef}.
\section{Expectation values}\label{expval}
In this section the time evolution of the expectation values is discussed in order to get a better understanding of the dynamics.
The time-dependence of the
expectation values is derived in the Heisenberg picture. Using the expressions in equation~\eqref{la-Gammaev} the following
calculations can be carried out without difficulty, however, the expressions appear a bit extensive due to the number of involved operators.
The Heisenberg operators for $\ON_1(t)$ resp. $\ON_2(t)$ read
\begin{align}\label{la-evN12}
\ON_1(t) &= \OU^{-1}(t)\ON_1\OU(t)= \OU_I^{-1}(t)\ON_1\OU_I(t)= \nonumber\\&\quad
\ON_1+\rmi\Big[\chi_t^{(1,0)}\OK_{1,0}-\chi_t^{(-1,0)}\OK_{-1,0}
+\chi_t^{(1,1)}\OK_{1,1}-\chi_t^{(-1,-1)}\OK_{-1,-1}\nonumber\\&\quad+
\chi_t^{(1,-1)}\OK_{1,-1}-\chi_t^{(-1,1)}\OK_{-1,1}\Big]\,,\\
\ON_2(t) &= \OU_I^{-1}(t)\ON_2\OU_I(t)= \ON_2+\rmi\Big[\chi_t^{(0,1)}\OK_{0,1}-\chi_t^{(0,-1)}\OK_{0,-1}\nonumber\\
&\quad+ \chi_t^{(1,1)}\OK_{1,1}-\chi_t^{(-1,-1)}\OK_{-1,-1}
-\chi_t^{(1,-1)}\OK_{1,-1}+\chi_t^{(-1,1)}\OK_{-1,1}\Big]\,.
\end{align}
Operators in the Heisenberg picture are indicated in the notation by the explicit time argument. The only differences between the operators $\ON_1(t)$ and $\ON_2(t)$ are the changed sign for the term with the operators $\OK_{1,-1}$ and $\OK_{-1,1}$ and the expressions $\chi_t^{(0,\pm1)}\OK_{0,\pm1}$ instead of $\chi_t^{(\pm1,0)}\OK_{\pm1,0}$.
More involved is the calculation of $N_1^2(t) = [N_1(t)]^2$ which can be obtained by squaring  equation~\eqref{la-evN12}:
\begin{align}
\ON_1^2(t) &= \ON_1^2 +2|\chi_t^{(1,0)}|^2+2|\chi_t^{(1,1)}|^2+2|\chi_t^{(1,-1)}|^2
-({\chi_t^{(1,0)2}}+2\chi_t^{(1,1)}\chi_t^{(1-1)})\OK_{2,0}\nonumber\\
&\quad+(\chi_t^{(-1,0)2}+2\chi_t^{(-1,-1)}\chi_t^{(-1,1)})\OK_{-2,0}
-\chi_t^{(1,1)2}\OK_{2,2}+\chi_t^{(-1,-1)2}\OK_{-2,-2}\nonumber\\&\quad
-\chi_t^{(1,-1)2}\OK_{2,-2}+\chi_t^{(-1,1)2}\OK_{-2,2}
+\rmi\Big[\chi_t^{(1,0)}\OJ_1-\chi_t^{(-1,0)}\OJA_1
+\chi_t^{(1,1)}\OJ_1\OK_{0,1}\nonumber\\&\quad-\chi_t^{(-1,-1)}\OJA_1\OK_{0,-1}
+\chi_t^{(1,-1)}\OJ_1\OK_{0,-1}-\chi_t^{(-1,1)}\OJA_1\OK_{0,1}\Big]
-2\chi_t^{(1,0)}\chi_t^{(1,1)}\OK_{2,1}\nonumber\\&\quad
-2\chi_t^{(-1,0)}\chi_t^{(-1,-1)}\OK_{-2,-1}
+2(\chi_t^{(1,0)}\chi_t^{(-1,-1)}+\chi_t^{(-1,0)}\chi_t^{(1,-1)})\OK_{0,-1} \nonumber\\
 &\quad
+2(\chi_t^{(-1,0)}\chi_t^{(1,1)}+\chi_t^{(1,0)}\chi_t^{(-1,1)})\OK_{0,1}
-2\chi_t^{(1,0)}\chi_t^{(1,-1)}\OK_{2,-1}
\nonumber\\&\quad
-2\chi_t^{(-1,0)}\chi_t^{(-1,1)}\OK_{-2,1}
+2\chi_t^{(1,1)}\chi_t^{(-1,1)}\OK_{0,2}
+2\chi_t^{(-1,-1)}\chi_t^{(1,-1)}\OK_{0,-2}\,.
\end{align}
Here, the operator $\OJ_j$ is the anti-commutator of $\ON_j$ and
$\OK_j$,
\begin{equation}
\OJ_j = [\ON_j,\OK_j]_+ =\ON_j\,\OK_j+\OK_j\ON_j\,.
\end{equation}
The expression for $\ON_2^2(t)$, obtained by squaring $\ON_2(t)$, has the same structure as $\ON_1^2(t)$. Instead of the functions $\chi_t^{(\pm1,0)}$ the functions $\chi_t^{(0,\pm1)}$ appear and, analogously, the indices of the operators $\OK_{u,v}$ are exchanged.
\subsection*{Time-independent case:}
Now we turn back to the time-independent case in order to analyse the Bloch oscillations in more detail.
The expectation values for the operators $\OK_{u,v}$, $\OJ_1\OK_2^v$ and $\OK_1^u\OJ_2$ can be decomposed as
\begin{align}\label{la-2dev}
\ev{\OK_{u,v}} &= K_{u,v} = |K_{u,v}|\,\rme^{\rmi\kappa_{u,v}}\\
\ev{\OJ_1\OK_2^v} &= J_{v} = |J_{v}|\,\rme^{\rmi\mu_{v}}\\
\ev{\OK_1^u\OJ_2} &= L_{u} = |L_{u}|\,\rme^{\rmi\nu_{u}}\,\,,
\end{align}
where the operator $\OK_1^u$ resp. $\OK_2^v$ corresponds to $\OK_{u,0}$ resp. $\OK_{0,v}$ in the alternative notation~\eqref{Kuvdef}. The expectation value $\ev{\OK_1^u\OJ_2}$ only appears in the expression for $\ev{\ON_2^2}_t$ (cf. equation~\eqref{expN22sep}). All expectation values including only the unitary
operators $\OK_{u,v}$ have modulus smaller than one
except for the eigenstates of these operators  where they have exactly modulus one.

In order to calculate the expectation values for the operators $\ON_1(t)$, $\ON_2(t)$ and $\ON_1^2(t)$ the functions $\chi_t^{(u,v)}$ are also decomposed into their moduli and phases,
\begin{equation}
\chi_t^{(u,v)} = |\chi_t^{(u,v)}|\,\rme^{-\rmi (uq+vr)\omm}\quad\text{and}\quad
\chi_t^{(-u,-v)} = |\chi_t^{(u,v)}|\,\rme^{+\rmi (uq+vr)\omm}\,.
\end{equation}
Then, the desired expressions for the expectation values read
\begin{align}\label{devN1}
\ev{\ON_1}_t &= \ev{\ON_1}_0
+2|\chi_t^{(1,0)}|\,|K_{1,0}|\sin\left(q\omm-\kappa_{1,0}\right)\nonumber\\&\quad
+2|\chi_t^{(1,1)}|\,|K_{1,1}|\sin\left((q+r)\omm-\kappa_{1,1}\right)\nonumber\\&\quad
+2|\chi_t^{(1,-1)}|\,|K_{1,-1}|\sin\left((q-r)\omm-\kappa_{1,-1}\right)\,,\\\label{devN2}
\ev{\ON_2}_t& = \ev{\ON_2}_0
+2|\chi_t^{(0,1)}|\,|K_{0,1}|\sin\left(r\omm-\kappa_{0,1}\right)\nonumber\\&\quad
+2|\chi_t^{(1,1)}|\,|K_{1,1}|\sin\left((q+r)\omm-\kappa_{1,1}\right)\nonumber\\&\quad
-2|\chi_t^{(1,-1)}|\,|K_{1,-1}|\sin\left((q-r)\omm-\kappa_{1,-1}\right)
\end{align}
and
\begin{align}\label{devN12}
\ev{\ON_1^2}_t &= \ev{\ON_1^2}_0
+2|\chi_t^{(1,0)}|^2\left(1-|K_{2,0}|
\cos\left(2q\omm-\kappa_{2,0}\right)\right) \nonumber\\&\quad
+2|\chi_t^{(1,1)}|^2\left(1-|K_{2,2}|
\cos\left(2(q+r)\omm-\kappa_{2,2}\right)\right)\nonumber\\&\quad
+2|\chi_t^{(1,-1)}|^2
\left(1-|K_{2,-2}|\cos\left(2(q-r)\omm-\kappa_{2,-2}\right)\right)\nonumber\\&\quad
+2|\chi_t^{(1,0)}|\,|J_{0}|\sin\left(q\omm-\mu_{0}\right)
+2|\chi_t^{(1,1)}|\,|J_{1}|\sin\left((q+r)\omm-\mu_{1}\right)\nonumber\\&\quad
+2|\chi_t^{(1,-1)}|\,|J_{-1}|\sin\left((q-r)\omm-\mu_{-1}\right)\nonumber\\&\quad
-4|\chi_t^{(1,0)}\chi_t^{(1,1)}|\,|K_{2,1}|\cos\left((2q+r)\omm-\kappa_{2,1}\right)\nonumber\\&\quad
+4|\chi_t^{(1,0)}\chi_t^{(1,1)}|K_{0,1}|\cos\left(r\omm-\kappa_{0,1}\right)\nonumber\\&\quad
-4|\chi_t^{(1,1)}\chi_t^{(1,-1)}|\,|K_{2,0}|\cos\left(2q\omm-\kappa_{2,0}\right)\nonumber\\&\quad
+4|\chi_t^{(1,0)}\chi_t^{(1,-1)}|\,|K_{0,1}|\cos\left(r\omm-\kappa_{0,1}\right)\nonumber\\&\quad
-4|\chi_t^{(1,0)}\chi_t^{(1,-1)}|\,|K_{2,-1}|\cos\left((2q-r)\omm-\kappa_{2,-1}\right)\nonumber\\&\quad
+4|\chi_t^{(1,1)}\chi_t^{(1,-1)}|\,|K_{0,2}|\cos\left(2r\omm-\kappa_{0,2}\right)\,.
\end{align}
It should be pointed out again that the functions $\chi_t^{(u,v)}$ depend only on system parameters ($g^{(u,v)}$ and $\vec{f}$) while $|K_{u,v}|$ and $\kappa_{u,v}$ resp. $|J_{\nu}|$ and $\mu_{\nu}$ are determined by the initial state.
As long as the field is {\it not} directed along the axes or the main
diagonals, i.~e. $q\!\neq\!0\!\neq\! r$, and $q+r\!\neq\! 0\!\neq\! q-r$, the
expectation values $\ev{\ON_1}_t$, $\ev{\ON_2}_t$, $\ev{\ON_1^2}_t$ and
$\ev{\ON_2^2}_t$ and therefore the widths $\Delta_{N_j}^2(t) =
\ev{\ON_j^2}_t-\ev{\ON_j}_t^2$, $j=1,2$, are {\it periodic} in time since all functions $\chi_t^{(u,v)}$ are periodic according to equation \eqref{chito}. Thus, the
results from the one-dimensional case concerning Bloch oscillations and
breathing modes remain valid: A wavepacket performs a two-dimensional Lissajous-like oscillation which shows no systematic dispersion, i.e. $\Delta_{N_j}^2(t)$ is bounded in time.

More interesting is the case that the
field is directed along one axis or a diagonal, i.~e.
\begin{equation}
(q,r) = (1,0),\,(0,1),\,(\pm1,\pm1),\,(\mp1,\pm1)\,.
\end{equation}
According to equation~\eqref{chito}, the function $\chi_t^{(r,-q)}$ grows linearly in
time, $\chi_t^{(r,-q)}= g^{(r,-q)}\,t$.
Since the argument $(uq+vr)\omega_Bt/2 $ of the related $\sin$-function in the expressions for the expectation values $\ev{\ON_j}_t$ (cf. equations~\eqref{devN1}, \eqref{devN2}) vanishes for all times $t$,
only the phase factor $\kappa_{r,-q}$ remains.
This has the consequence that the term containing  $\chi_t^{(q,-r)}$ in the equation for $\ev{\ON_1}_t$ resp. $\ev{\ON_2}_t$ increases linearly in time as far as the corresponding phase factor $\kappa_{u,v}$ does not vanish either.\\
Thus $(\ev{\ON_1}_t,\ev{\ON_2}_t)$ performs an oscillatory motion (due to the remaining terms) with a superimposed linear motion orthogonal to the field direction.\\
For $(q,r) = (1,0)$ it is clear that the linear motion is orthogonal to the field, since the linearly growing function $|\chi_t^{(0,1)}| = |\chi_t^{(0,-1)}|$ appears only in the expression for $\ev{\ON_2}_t$. The same argument holds for $(q,r) = (0,1)$. If the field is directed along one of the main diagonals the motion is also  orthogonal to the field as can be checked directly.
In general, the velocity of the directed motion is determined via equation~\eqref{la-Gammaev} by the commutators~\eqref{Kuvcom} and can be written as
\begin{equation} \label{veldm}
\vec{v}=2g^{(r,-q)}\,|K_{r,-q}|\sin(\kappa_{r,-q})\,\begin{pmatrix}
		\epsilon_{r,-q,1}\\\epsilon_{r,-q,2}
           \end{pmatrix}
	 =2g^{(r,-q)}\,|K_{r,-q}|\sin(\kappa_{r,-q})\, \begin{pmatrix}
	 r\\-q
	   \end{pmatrix}\,.
\end{equation}

Independent of the initial state, $\ev{N_j^2}_t$ grows quadratically in time due to the second up to
fourth term in equation~\eqref{devN12}. The contribution from the other  terms is at most linear. The case that the factors cancel each other can be excluded because this requires $|K_{u,v}| =1$ which is only fulfilled if the initial state of the system is an eigenstate of the shift operator $\OK_{u,v}$. But these states are delocalised in a basis of Wannier-states and can therefore be excluded.
With $\ev{N_j^2}$ also the variance $\Delta_{N_j}^2(t) =\ev{\ON_j^2}_t-\ev{\ON_j}_t^2$, $j=1,2\,,$ grows quadratically in time. Only for states with special initial phases $\kappa_{u,v}$ and amplitudes $|K_{u,v}|$ a localisation effect can be achieved. But this is no systematic effect, i.e. not depending on the field and the coupling constants, but only caused by a specially prepared initial state. In section~\ref{sec-basis}, this behaviour is analysed in more detail for a Gaussian initial state.
\subsection*{Separable Case:}
Before we conclude this section, a brief discussion of the separable
case, where only
couplings along the axes are taken into account, allows further
insight into the dynamics. Here, we also assume that the field is time-independent. For the separable system with $\Delta_{3}=0$ in the Hamiltonian~\eqref{la-2dHam}, the expressions for the expectation values, equations~\eqref{devN1} and \eqref{devN2}, simplify
drastically:
\begin{align}\label{la-2devNsep}
\ev{\ON_1}_t& \stackrel{\phantom{q\neq0}}{=}
\ev{\ON_1}_0+2|\chi_t^{(1,0)}|\,|K_{1,0}|\sin\left(q\omm-\kappa_{1,0}\right)\nonumber\\
&\stackrel{q\neq0}{=}\ev{\ON_1}_0+\frac{\Delta_1}{q\omega_B} |K_{1,0}|
\sin\left(q\omm\right)\sin\left(q\omm-\kappa_{1,0}\right)\nonumber\\
&\stackrel{\phantom{q\neq0}}{=}\ev{\ON_1}_0+\frac{\Delta_{1}}{2q\omega_B} |K_{1,0}|
\left[ \cos\left(q\omega_Bt-\kappa_{1,0}\right)-\cos(\kappa_{1,0})\right]\,,\\
\ev{\ON_2}_t& \stackrel{\phantom{q\neq0}}{=} \ev{\ON_2}_0
+2|\chi_t^{(0,1)}|\,|K_{0,1}|\sin\left(r\omm-\kappa_{0,1}\right)\nonumber\\
&\stackrel{r\neq0}{=}\ev{\ON_2}_0+\frac{\Delta_{2}}{2r\omega_B} |K_{0,1}|
\left[ \cos\left(r\omega_Bt-\kappa_{0,1}\right)-\cos (\kappa_{0,1})\right]\,.
\end{align}
If the field is not directed along one of the axes, the result is a Lissajous-like oscillation of the wavepacket depending
on the field direction $(q,r)$ and both initial phases $\kappa_{1,0}$
and $\kappa_{0,1}$\, (and not only the phase difference $\Delta\phi =
\kappa_{1,0} - \kappa_{0,1}$).
According to equation~\eqref{devN2}, the expectation values for
$\ON_j^2(t)$ read
\begin{align}
\ev{\ON_1^2}_t &= \ev{\ON_1^2}_0
+2|\chi_t^{(1,0)}|^2\left[1-|K_{2,0}| \cos\left(2q\omm-\kappa_{2,0}\right)\right]\nonumber\\
&\quad+2|\chi_t^{(1,0)}|\,|J_{0}|\sin\left(q\omm-\mu_{0}\right)\,,
\end{align}
\begin{align}\label{expN22sep}
\ev{\ON_2^2}_t =& \ev{\ON_2^2}_0
+2|\chi_t^{(0,1)}|^2\left[1-|K_{0,2}| \cos\left(2r\omm-\kappa_{0,2}\right)\right]\nonumber\\
&\quad+2|\chi_t^{(0,1)}|\,|L_{0}|\sin\left(r\omm-\nu_{0}\right)\,.
\end{align}
These are periodic functions of time if the field direction is not
along one of the axes.
The variances $\Delta_{\ON_j}^2(t)$ can be calculated directly:
\begin{align}
\Delta_{\ON_1}^2(t) &=\Delta_{\ON_1}^2(0)\nonumber\\&+2|\chi_t^{(1,0)}|^2\Bigl[1-|K_{2,0}| \cos\left(2q\omm-\kappa_{2,0}\right)-2|K_{1,0}|^2\sin^2\left(q\omm-\kappa_{1,0}\right)\bigr]\nonumber\\
 &+2|\chi_t^{(1,0)}|\left[|J_{0}|\sin\left(q\omm-\mu_{0}\right)-|K_{1,0}|\ev{\ON_1}_0 \sin\left(q\omm-\mu_{1,0} \right)\right]\,,
\end{align}
\begin{align}
\Delta_{\ON_2}^2(t) &= \Delta_{\ON_2}^2(0)\nonumber\\&
+2|\chi_t^{(0,1)}|^2\left[1-|K_{0,2}| \cos\left(2r\omm-\kappa_{0,2}\right)-2|K_{0,1}|^2\sin^2\left(r\omm-\kappa_{0,1}\right)\right]\nonumber\\
 &+2|\chi_t^{(0,1)}|\left[|L_{0}|\sin\left(q\omm-\nu_{0}\right)-|K_{0,1}|\ev{\ON_2}_0 \sin\left(r\omm-\mu_{0} \right)\right]\,.
\end{align}
For a field direction along one axis, e.~g. the first, a wavepacket shows dispersion in the
direction of the second axis due to $\chi_t^{(0,1)}=g^{(0,1)}t$. Besides, $\ev{\ON_2}_t$ can increase also
linearly in time, if the initial state provides a phase $\kappa_{0,1}\neq n\pi$, $n\in\Z$.

Concluding, for a field directed along an axis or the main diagonal the
Bloch oscillation of the centre of mass of a wavepacket is superposed by a directed motion orthogonal to the field. Only for special initial conditions this directed motion is not present and one finds strictly periodic oscillations. Nevertheless, the projection of the motion onto the field direction is always periodic.
For other field directions the Bloch oscillations still remain, since the couplings orthogonal to the field, $g^{(r,-q)}$, are not included into this tight-binding model. But,
extending the tight-binding approximation by taking into account the coupling to more directions leads to a superposed directed motion also in these cases, as will be seen in the next section.
\section{Single-band models}\label{sbm}
As in the one-dimensional case (cf. \S 9 in \cite{03TBalg}) one can add an arbitrary number of additional shift
operators
\begin{equation}
\OK_1\OK_1\OK_2,\;\OK_1\OK_2\OK_2,\:\OK_1\OK_1\OK_1\OK_2,\;
\OK_1\OK_1\OKA_2,\;...
\end{equation}
to the algebra.
Physically
these operators couple adjacent sites in the directions given by their indices ($\OK_{u,v}$ with $u$ and $v$ coprime, is the coupling in direction $(u,v)$) or describe the coupling to further sites ($\OK_{nu,nv}$ with $n\in\N$ and $u$, $v$ coprime is the coupling to the $n$th site in direction $(u,v)$). Generalising expression~\eqref{lauv}, the Lie-algebra can be written as
\begin{equation}
\LL = \left\{\ON_1,\ON_2,\OK_{u,v}\Big|\,(u,v)=\M \right\}\quad\text{with}\;\M = \Z^2\setminus\{(0,0)\}\,.
\end{equation}
The corresponding Hamiltonian is a direct extension of the tight-binding Hamiltonian and reads
\begin{equation}
H = \sum_{(u,v)\in\M}g^{(u,v)}\OK_{u,v}+f_1\ON_1+f_2\ON_2\quad\text{with}\quad g^{(u,v)}=g^{(-u,-v)}\,.
\end{equation}
The evolution operator can be calculated in the same way as in section~\ref{teo}, $\OU(t) = \OU_S(t)\OU_I(t)$ with $\OU_S(t)$ defined in equation~\eqref{USdef} and
\begin{equation}
\OU_I(t) = \exp\biggl\{-\rmi\sum_{(u,v)\in\M}\chi_t^{(u,v)}\OK_{u,v}\biggr\}\,.
\end{equation}
Instead of only eight functions $\chi_t^{(u,v)}$ in the tight-binding approximation, one gets infinitely many functions. Therefore, for every
rational field direction $(q,r)$ there exist pairs $(nr,-nq)$, $n\in\Z$, so that  $\chi_t^{(u,v)} = g^{(u,v)}\,t$ with the same consequences as in the tight-binding approximation.
Especially, the velocity of the directed motion is given by equation~\eqref{veldm}.
But, one has to keep in mind that for realistic potentials, as investigated for example in \cite{04bloch2d}, the factors $g^{(u,v)}$ decrease
exponentially with increasing $|u|+|v|$ so that the directed motion and the dispersion are very small compared to the oscillation.
\section{Case study of a Gaussian initial distribution in a Wannier basis}\label{sec-basis}
In this section, the results from section~\ref{expval} will be
applied to an explicit representation of the initial wavepacket in a
basis of Wannier functions,
\begin{equation}
\ket{\psi(t)} = \sum_{m,n}c_{m,n}(t)\ket{m,n}\,.
\end{equation}

\begin{figure}[htb]
\begin{center}
\includegraphics[width=6.5cm,  angle=0]{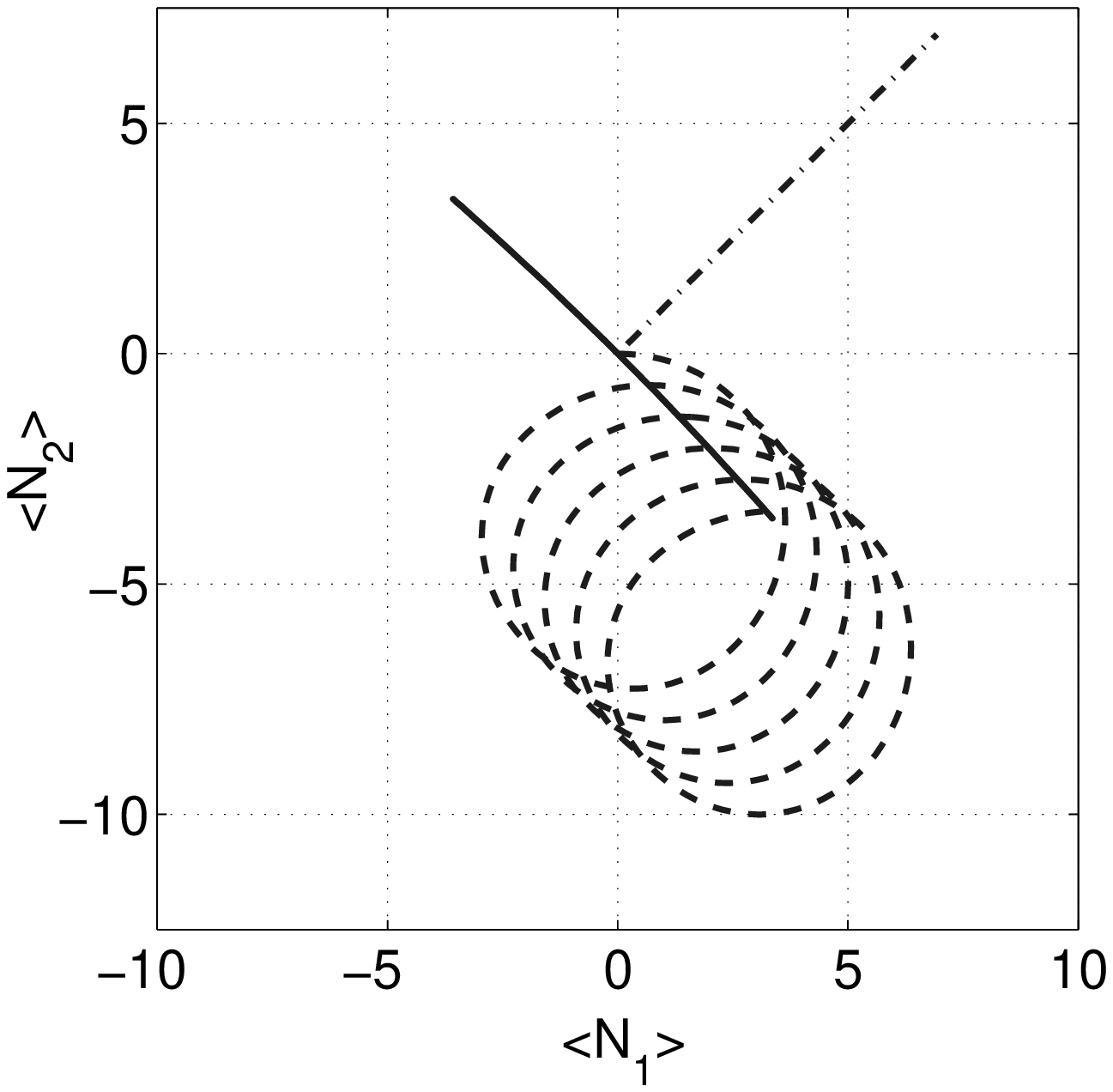}
\includegraphics[width=6.5cm,  angle=0]{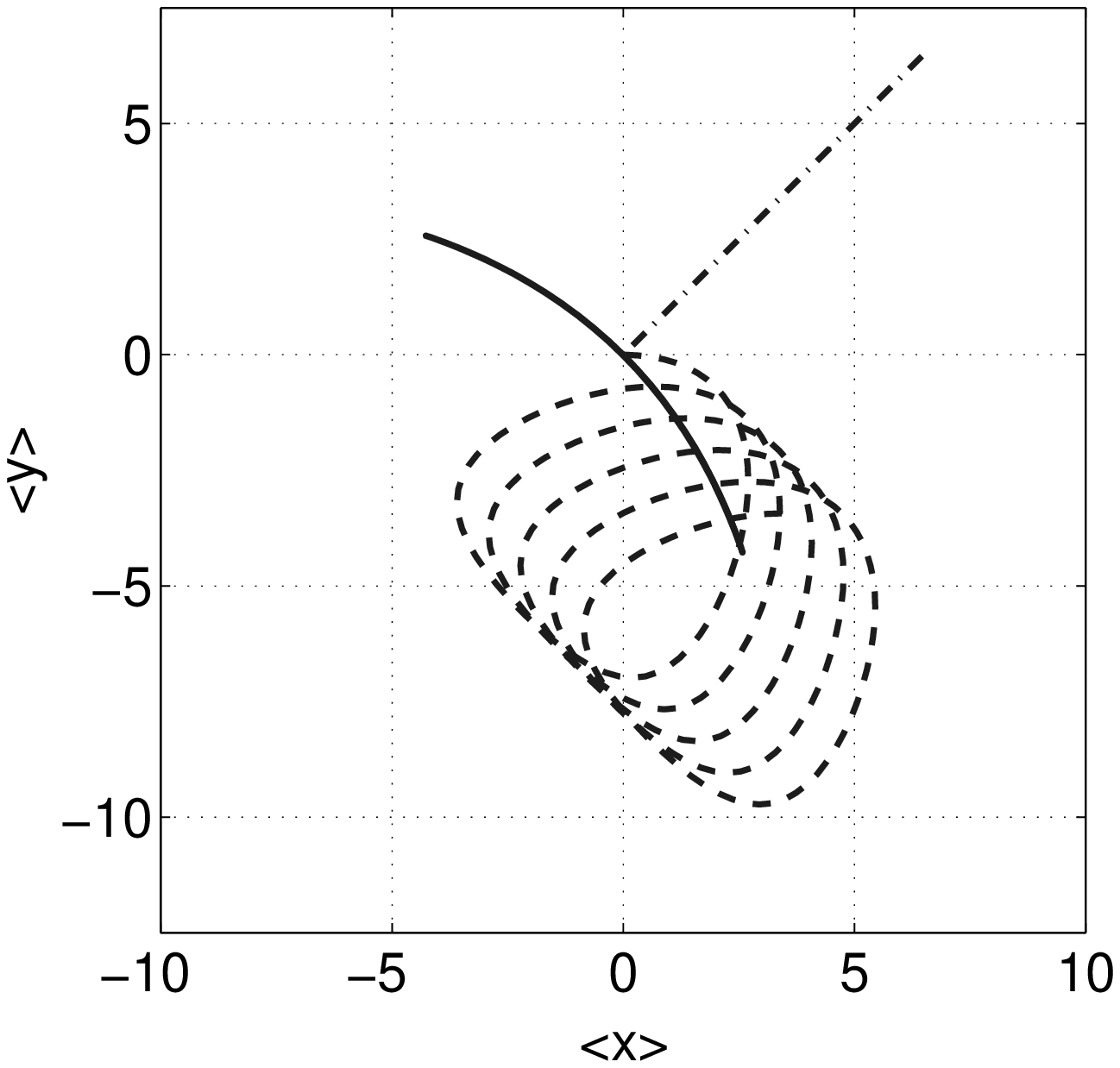}
\end{center}
\caption{\label{blochosz}Left-hand: Time dependence of the expectation values
$(\ev{N_1}_t,\ev{N_2}_t)$ for a field direction $q\!=\!r\!=\!1$ and three
different initial conditions. The dashed line shows the trajectory for
$(k_1,k_2)=(\pi/2,0)$, the
dashed-dotted line for $(k_1,k_2)=(\pi,\pi)$ and the solid
line for $(k_1,k_2)=(\pi/2,-\pi/2)$\,.
Right-hand: Trajectories $(\ev{x},\ev{y})$ obtained by a direct integration of the time-dependent Schr\"odinger equation and the Hamiltonian~\eqref{WSH} with potential~\eqref{potxy} for the same parameters and initial conditions as on the left-hand side.}
\end{figure}

The expectation values can be calculated using
expressions \eqref{devN1}, \eqref{devN2} and \eqref{devN12}. The expectation values $ \ev{\OK_{u,v}} =
K_{u,v}$ can be evaluated as
\begin{equation}\label{kuvgauss}
\ev{\OK_{u,v}} = \scal{\psi}{\OK_1^u\OK_2^v|\psi} =
\sum_{m,n}c^*_{m,n}c_{m+u,n+v}
\end{equation}
with $c_{m,n} = c_{m,n}(0)$ (cf. equation~\eqref{la-2dOpdef2}).
In the following, we will assume that the initial distribution is
Gaussian, i.~e.
\begin{equation}
c_{m,n} = \alpha\exp\left\{-\frac{m^2+n^2}{4\sigma^2}+\rmi(k_1m+k_2n)\right\}
\end{equation}
with normalisation constant $\alpha\in\R$.
Then, the expectation value \eqref{kuvgauss}  can be written as
\begin{eqnarray}\nonumber
\ev{K_{u,v}} &=\alpha^2\rme^{\rmi(k_1u+k_2v)}
\sum_{m,n}\rme^{-(m^2+n^2+(m+u)^2+(n+v)^2)/(4\sigma^2)}\\
 &=|K_{u,v}|\,\rme^{\rmi(k_1u+k_2v)}\,\label{GKuv}
\end{eqnarray}
and the phase factor in equation~\eqref{la-2dev} is given by
\begin{equation}\label{wbkappa}
\kappa_{u,v} = k_1u+k_2v \,.
\end{equation}
Reconsidering the expressions for $\ev{\ON_1}_t$ and $\ev{\ON_2}_t$ in equations \eqref{devN1} and \eqref{devN2}, the
behaviour of the wavepacket depends crucially on this phase
$\kappa_{u,v}$.
The function $\chi_t^{(r,-q)}$, which describes the coupling strength in the direction orthogonal to the field, grows linearly in time according to equation~\eqref{chito}.
Then, depending on the argument of the related sin-function, $\sin(\kappa_{r,-q})$, this linearly growing term contributes to the expectation value or is cancelled by $\sin(\kappa_{r,-q})\!=\!0$. If $\kappa_{r,-q}$ is an integer multiple of $\pi$,  this term does not contribute and the expectation value is periodic in time.
That means that for a given field direction $(q,r)$ one has to
make sure that the condition
\begin{equation}
\kappa_{r,-q} = k_1r-k_2q = n\pi\quad\text{with}\quad n\in\Z
\end{equation}
is fulfilled to get a periodic centre of mass motion.
If this condition is not fulfilled, the periodic motion is superposed by a directed motion orthogonal to the field.
The left-hand side of figure~\ref{blochosz} illustrates this behaviour. Displayed are the expectation values $\ev{\ON_1}$ and $\ev{\ON_2}$ as a function of time which are obtained by a direct numerical integration of the time-dependent Schr\"odinger equation for the Hamiltonian~\eqref{la-2dHam} and parameter values $\hbar=2.81 / 2\pi$, $F = 0.1$, $\Delta_1 = \Delta_2 = 1/4$, $ \Delta_{3}=0.008$.

Independent of the phase \eqref{wbkappa} of the initial wavepacket, the width of the wavepacket increases quadratically in time. Considering for example equation \eqref{devN12} and a field along the second axis or one of the main diagonals this can be seen in the following way. The relevant term for the quadratical dispersion has the form
\begin{equation}
2|\chi_t^{(r,-q)}|^2\left(1-|K_{2r,-2q}|
\cos\left(\kappa_{2r,-2q}\right)\right)\,.
\end{equation}
Together with the relevant term for $\ev{N_1}_t$,
\begin{equation}
2|\chi_t^{(r,-q)}|\,|K_{r,-q}|\sin\left(\kappa_{r,-q}\right)
\end{equation}
one gets for the width of the wavepacket
\begin{align}\nonumber
\Delta^2_{\ON_1}(t) &= \ev{\ON_1^2}_t-\ev{\ON_1}_t^2 \\&\approx2|\chi_t^{(r,-q)}|^2\left[1-|K_{2r,-2q}|
\cos\left(\kappa_{2r,-2q}\right)-2|K_{r,-q}|^2\sin^2\left(\kappa_{r,-q}\right)\right]\,.\label{deltaGN12}
\end{align}
The phase factor $\kappa_{2r,-2q}$ can be written as $2\kappa_{r,-q}$ using equation~\eqref{wbkappa} so that the following inequality holds
\begin{align}
&1-|K_{2r,-2q}|
\cos\left(\kappa_{2r,-2q}\right)-2|K_{r,-q}|^2\sin^2\left(\kappa_{r,-q}\right)\nonumber\\&
=1-(2|K_{r,-q}|^2-|K_{2r,-2q}|)\sin^2\left(\kappa_{r,-q}\right)-|K_{2r,-2q}|
\cos^2\left(\kappa_{r,-q}\right)>0
\,.
\end{align}
Here we also used that $|K_{u,v}|<1$ and $0<2|K_{u,v}|^2-|K_{2u,2v}|<1$ for a localised initial state. The second relation can be shown by using equation~\eqref{GKuv} and the relations for the $\vartheta$-function.
Inserting this inequality into equation~\eqref{deltaGN12} shows that the quadratically increasing term $|\chi_t^{(r,-q)}|^2$ contributes to $\Delta^2_{\ON_1}(t)$ for every value of $K_{u,v}$ resp. $\kappa_{u,v}$. Therefore, every Gaussian initial wavepacket shows a quadratic dispersion.

The tight-binding results can be compared with a direct integration of the time-dependent Schr\"odinger equation for the Hamiltonian~\eqref{WSH}. The right-hand side of figure~\ref{blochosz} shows the trajectories of the expectation values $(\ev{x}_t,\ev{y}_t)$ for a Gaussian wavepacket
\begin{equation}
\psi(x,y)=(2\pi \sigma^2)^{-1/2}\,\rme^{-(x^2+y^2)/4\sigma^2
+\rmi\, k_1\,x+\rmi\, k_2\,y}
\end{equation}
with $\sigma=10\pi$ which is projected onto the first Bloch-band:
\begin{equation}
\psi_1(x,y) = \scal{x,y}{\psi_1} = \sum_{m,n}\scal{m,n}{\psi}\,\scal{x,y}{m,n}
\end{equation}
for three initial momenta $(k_1,k_2)=(\pi/2,0),\;(\pi,\pi),\;(\pi/2,-\pi/2)$. The system parameters are $\hbar=2.81/2\pi$ and $F=0.1$ in scaled units. The potential
\begin{equation}\label{potxy}
V(x,y) = \cos(2\pi x) + \cos(2\pi y) - \cos(2 \pi x) \cos(2 \pi y)
\end{equation}
generates a lowest Bloch band in the energy dispersion relation with width $\Delta_1 = \Delta_2 = 1/4$ along the axes and a width $\Delta_3 = 0.008$ along the main diagonals, so that the results can be directly compared with the tight-binding calculations on the left-hand side of figure~\ref{blochosz}. The small distortion compared to the tight-binding results can be explained by the couplings to further sites in the single-band model.
\section{Concluding remarks}\label{s-conclusions}
In the present paper we have investigated Bloch oscillations in a two-dimensional tight-binding resp. single-band model. Using an algebraic approach, the evolution operator has been expressed in a product form for an arbitrary time dependent field. It has been shown analytically for the expectation values that an initial state in general does not perform a strictly periodic centre of mass Bloch oscillations in the single-band approximation. While the projection of this motion onto the field direction still oscillates, the projection orthogonal to the field additionally shows a directed motion. Only for special conditions that depend on the initial wavepacket and not on the system parameters the strict periodicity of the Bloch oscillations can be restored.
The width of a wavepacket behaves similar: It shows a systematic quadratical dispersion orthogonal to the field that can only be suppressed for special initial wavepackets.

Finally, the similarity to the one-dimensional tight-binding system with an additional harmonic driving, $F(t) = F_0+F_1\cos\omega t$ should be emphasised. For an integer ratio between the Bloch frequency and the driving frequency, $\omega_B = q \,\omega$ ($q\in\N$) a wavepacket shows a systematic dispersion. But, in contrast to the two-dimensional dispersion , the dispersion can be suppressed by adjusting the field strength $F_1$ and the driving frequency $\omega$ (dynamical localisation) \cite{Dunl86,Grif98,Holt96}. \\
Therefore, the two-dimensional tight-binding system with an additional time-periodic driving can be expected to show a multitude of interesting phenomena which have to be analysed in more detail.
\section*{Acknowledgements}
Support from the Deutsche Forschungsgemeinschaft
via the Graduiertenkolleg  ``Nichtlineare Optik und Ultrakurzzeitphysik''
as well as from the Volkswagen Foundation is gratefully acknowledged.

\end{document}